# A Simple Efficient Method for Obtaining the Binding Energy of Lithium Nucleus under the Hulthén and Inversely Quadratic Yukawa Potentials

Nasrin. Salehi[*1], Mahsa. Ghazvini[2]

[1]Department of Basic Sciences, Shahrood Branch, Islamic Azad University, Shahrood, Iran
[2]Department of physics, Shahrood University, Shahrood, Iran

**Abstract**

In this paper, the binding energy of Lithium nucleus in a nonrelativistic model is obtained for the Hulthén and the Inversely Quadratic Yukawa Potential. In order to that, we used the concept of supersymmetry to solving the Schrödinger equation exact analytically. These potentials, due to their physical interpretations, are of interest within many areas of theoretical physics. The results of our model for all calculations show that the ground state binding energy of Lithium nucleus with these potentials are very close to the ones obtained in experiments.

## 1. Introduction

The subject of binding energy has received great theoretical attention because of many attractive features they exhibit in a large class of combinations and different methods have been used. Among the different forms of physical potentials which appear in the Hamiltonian, those with spherical symmetry have received great attention within the recent years because of their wide applications (de Souza and Almeida, 2000; Roy and Roy, 2002; Alhaidari, 2002; Serra and Lipparini, 1997; Li et al., 2003; Dong et al., 1999). On the other hand, since the study of many physical systems corresponds to study an N-dimensional problem, many attempts have been made to analyze N-dimensional spaces (Dong, 2000; Arda and Sever, 2010; Milanovic and Ikovic, 1999; Cardoso and Alvarez- Nodarse, 2003; Coelho and Amaral, 2002; Chen, 2005; Kalnin et al., 2002; Oyewumi and Ogunsola, 2004). Many efforts is done to solve Schrödinger equation for 2 and 3 particles by using several methods such as NU method, Supersymmetry method(SUSY), and Anzast method with different potential [1-6]. Same work is done for other equations (for example Dirac equation, Klein-Gordon equation, ...) [7].
In the present work, we consider lithium nucleus as a 6-dimensional system, obtain the exact solution of the Schrödinger equation with the central potentials by using SUSY method.
In sections (2) and (3) we review Yukawa Potential and Hulthén Potential. In next section, Supersymmetry Method is reviewed. Then we study an analytical solution to the Schrödinger equation for 6-body system and we report the numerical results. Section (6) includes summery on the paper and conclusions.

## 2. The Inversely Quadratic Yukawa Potential

The Inversely Quadratic Yukawa (IQY) potential takes the following form:

$$V = \frac{V_0 e^{-2ar}}{r^2} \tag{1}$$

where $a$ is the screening parameter and $V_0$ is the depth of the potential. A form of Yukawa potential has been earlier used by Taseli [8] in obtaining modified Laguerre basis for hydrogen-like systems. Also, Kermode *et al.* [9] have used different forms of the Yukawa potential to obtain the effective range functions. But, not much has been done in solving the IQY potential.

## 3. The Hulthén Potential

The Hulthén potential has the following form:

$$V(r) = -V_0 \frac{e^{-\delta r}}{1 - e^{-\delta r}} \tag{2}$$

where $\delta$ is the screening parameter which is used for determining the range of the Hulthén potential. The parameter $V_0$ represents $\delta Z e^2$, where $Ze$ is the charge of the nucleon [10, 21-23]. The intensity of the Hulthén potential is denoted by $V_0$ under the condition of $\delta > 0$. The Hulthén potential is one of the important short-range potential which behaves like a Coulomb potential for small values of r and decreases exponentially for large values of r. The Hulthén potential has received extensive study in both relativistic and non-relativistic quantum mechanics [12, 19]. This potential has been used in several branches of physics and it's discrete and continuum states have been studied by a variety of techniques such as the formalism of supersymmetric quantum mechanics within the framework of the variational method [11] and the NU method. There is a wealth of literature on the use of the Hulthén potential as an approximation of the interaction potential in a number of area in physics such as nuclear and particle physics [13], atomic physics [14,15], solid state physics [16] and chemical physics [17]. Unfortunately, quantum mechanical equations with the Hulthén potential can be solved analytically only for states with zero-angular momentum [12, 18]. Recently, some interesting research papers have appeared to study the l-state solutions of the relativistic Klein-Gordon equation with the Hulthén potential [19, 20].

## 4. Supersymmetry Method

We start noticing that we know the ground state function of a 1dimension problem; we can find also the potential, up to a constant [6]. Taking the ground state energy to be zero, from the TISE we have:

$$H_1 \Psi_0(x) = -\frac{\hbar^2}{2m} \frac{d^2 \Psi}{dx^2} + V_1(x) \Psi_0 = 0 \tag{3}$$

$$V_1 = \frac{\hbar^2}{2m} \frac{\Psi_0''(x)}{\Psi_0(x)} \tag{4}$$

We can try to factorize the Hamiltonian with the ansatz:

$$H_1 = A^+A = H - E_0 \tag{5}$$

where:

$$A = \frac{\hbar}{\sqrt{2m}} \frac{d}{dx} + W(x), A^+ = -\frac{\hbar}{\sqrt{2m}} \frac{d}{dx} + W(x) \tag{6}$$

and $W(x)$ is called the superpotential and can be written in terms of the ground state as:

$$W(x) = -\frac{\hbar}{\sqrt{2m}} \frac{\Psi_0'}{\sqrt{\Psi_0}} \tag{7}$$

Writing $V_1$ in terms of $W(x)$ we obtain the Ricatti equation:

$$V_1(x) = W^2(x) - \frac{\hbar}{\sqrt{2m}} W'(x) \tag{8}$$

So now we can build up a SUSY theory searching for the SUSY partner Hamiltonian associated with $H_1$, namely $H_2 = AA^+$. This second Hamiltonian corresponds to a new potential:

$$H_2 = -\frac{\hbar^2}{2m} \frac{d^2}{dx^2} + V_2(x), V_2(x) = W^2(x) + \frac{\hbar}{\sqrt{2m}} W'(x) \tag{9}$$

## 5. Exact Analytical Solution of the Schrödinger Equation for 6-Body System with Certain Potentials

The Schrödinger equation in D-dimensions is

$$\frac{-\hbar^2}{2\mu}\left(\frac{d^2}{dr^2} + \frac{D-1}{r}\frac{d}{dr} - \frac{l(l+D-2)}{r^2}\right)R_{n,l}(r) + V(r)R_{n,l}(r) = E_{n,l}R_{n,l}(r), \quad D = 3N-3 \tag{10}$$

$N$ is the number of particle and $l$ is the angular momenta [24].

Here, consider the Schrödinger equation for 6-body system with a potential $V(r)$ that depends only on the distance $r$ from the origin:

$$\frac{-\hbar^2}{2\mu}\left(\frac{d^2}{dr^2} + \frac{14}{r}\frac{d}{dr} - \frac{l(l+13)}{r^2}\right)R_{n,l}(r) + V(r)R_{n,l}(r) = E_{n,l}R_{n,l}(r) \tag{11}$$

$$\left(\frac{d^2}{dr^2} + \frac{14}{r}\frac{d}{dr} - \frac{l(l+13)}{r^2}\right)R_{n,l}(r) + \frac{2\mu}{\hbar^2}\left(E_{n,l} - V(r)\right)R_{n,l}(r) = 0 \tag{12}$$

By applying $U_{n,l} = R_{n,l} r^{\frac{D-1}{2}} = R_{n,l} r^7$, we can write:

$$\frac{dR_{n,l}}{dr} = \frac{dU_{n,l}}{dr}r^{-7} - 7r^{-8}U_{n,l} \qquad \frac{d^2R_{n,l}}{dr^2} = \frac{d^2U_{n,l}}{dr^2}r^{-7} - 14r^{-8}\frac{dU_{n,l}}{dr} + 56r^{-9}U_{n,l} \qquad (13)$$

By substituting Eq. (13) in Eq. (12), we found the following form for Eq. (12):

$$\frac{d^2U_{n,l}}{dr^2}r^{-7} - 42r^{-9}U_{n,l} - l(l+13)r^{-9}U_{n,l} + \frac{2\mu}{\hbar^2}\left(E_{n,l} - V(r)\right)U_{n,l}r^{-7} = 0 \qquad (14)$$

By some summarizing, Eq. (14) changes to:

$$\frac{d^2U_{n,l}}{dr} + \frac{2\mu}{\hbar^2}\left(E_{n,l} - V(r) - \frac{\hbar^2(l+6)(l+7)}{2\mu r^2}\right)U_{n,l} = 0 \qquad (15)$$

where $l$ is the angular momentum quantum number and $\mu$ is the reduced mass.

It is suitable to introduce the IQY potential and use the Taylor expansion. So the potential takes the form:

$$V = \frac{V_0 e^{-2ar}}{r^2} = V_0\frac{(1 - 2ar + \frac{4a^2r^2}{2})}{r^2} = \frac{V_0}{r^2} - \frac{2aV_0}{r} + 2a^2V_0 \qquad (16)$$

By putting Eq. (16) into Eq. (15), the Schrödinger equation changes to:

$$\frac{d^2U_{n,l}(r)}{dr^2} + \frac{2\mu}{\hbar^2}\left(E_{n,l} - 2a^2V_0 + \frac{2aV_0}{r} - \frac{(V_0 + \Omega)}{r^2}\right)U_{n,l}(r) = 0 \qquad (17)$$

where $\Omega$ is defined as $\Omega = \frac{\hbar^2(l+6)(l+7)}{2\mu}$. By choosing $\varepsilon_{n,l} = \frac{2\mu}{\hbar^2}(E_{n,l} - 2a^2 V_0)$, $\beta = \frac{4\mu}{\hbar^2}V_0 a$,

$\gamma = \frac{2\mu}{\hbar^2}(V_0 + \Omega)$, Eq. (17) changes to:

$$\frac{d^2U_{n,l}(r)}{dr^2} + \left(\varepsilon_{n,l} + \frac{\beta}{r} - \frac{\gamma}{r^2}\right)U_{n,l}(r) = 0 \qquad (18)$$

In Supersymmetric Quantom Mechanics, the superpotential is defined as:

$$W_1 = -\frac{\hbar}{\sqrt{2\mu}}\left(A + \frac{B}{r}\right) \qquad (19)$$

Substituting this superpotential into Riccati equation that has the form of:

$$W_1^2(x) - \frac{\hbar}{\sqrt{2\mu}}W_1'(x) = \frac{2\mu}{\hbar^2}\left(V_1(x) - E_0^{(1)}\right) \qquad (20)$$

Then we reach to:

$$\left(A^2 + \frac{B^2}{r^2} + \frac{2AB}{r} - \frac{B}{r^2}\right) = \left(-\varepsilon_{n,l} - \frac{\beta}{r} + \frac{\gamma}{r^2}\right) \tag{21}$$

By doing some calculations, we can get $A^2 = -\varepsilon_{n,l}$, $2AB = -\beta$, $B^2 - B = \gamma$ and the ground state binding energy for Lithium nucleus is given as following:

$$E_{n,l} = \frac{-\hbar^2}{2\mu}\left(\frac{-\beta}{1 \pm \sqrt{1+4\gamma}}\right)^2 + V_0 a \tag{22}$$

By using Eq. (23) from SUSY method:

$$\psi_0^1(r) = N_0 \exp\left(-\frac{\sqrt{2\mu}}{\hbar}\int W(r')\,dr'\right) \tag{23}$$

the ground state normalized eigenfunctions are given as:

$$\psi_0^1(r) = N_0 \exp[Ar + B\ln r] \tag{24}$$

In Table 1. the fitted values of parameters of the ground state binding energy equations for IQY potential are given.

Table 1. The fitted values of parameters of the ground state binding energy equations for IQY potential, coloumn B.E (our model) contains our calculation and the column B.E (experiment) contains the experimental data.

| $a$ (fm$^{-1}$) | $V_0$ (MeV.fm$^2$) | B.E (MeV) | B.E (experiment) [26] |
|---|---|---|---|
| 0.55 | 50 | 30.22 | 31.995 |
| 0.50 | 50 | 24.97 | 31.995 |
| 0.60 | 45 | 32.37 | 31.995 |
| 0.59 | 45 | 31.30 | 31.995 |
| 0.57 | 47 | 31.59 | 31.995 |
| 0.59 | 50 | 34.77 | 31.995 |
| 0.58 | 48 | 32.26 | 31.995 |
| 0.58 | 47.5 | 31.93 | 31.995 |
| 0.58 | 47 | 31.59 | 31.995 |

From Table 1, it can be seen that for $a = 0.58$(fm$^{-1}$) and $V_0 = 47.5$(MeV), the calculated ground state binding energy (31.93 MeV) has a good agreement with the experimental data.

As we know from SUSY, the potential is determined as:

$$V_{\pm} = W^2 \pm \frac{\hbar}{\sqrt{2\mu}} \frac{dW}{dr} = \frac{\hbar^2}{2\mu}\left[A^2 + \frac{B^2}{r^2} + \frac{2AB}{r^2} \pm \frac{B}{r^2}\right] \qquad (25)$$

We obtain $A = \frac{-\beta}{2(B^2 - \gamma)}$ form Eq. (21). By substituting $A$ parameter into Eq. (25) we reach to:

$$V_{\pm} = \frac{\hbar^2}{2\mu}\left[(\frac{-\beta}{2(B^2-\gamma)})^2 + \frac{B^2}{r^2} + \frac{2\left(\frac{-\beta}{2(B^2-\gamma)}\right)B}{r^2} \pm \frac{B}{r^2}\right] \qquad (26)$$

The shape invariance concept that was introduced by Gendenshtein is [28]:

$$V_+(a_0, r) = V_-(a_1, r) + R(a_1) \qquad (27)$$

where $a_1$ is a function of $a_0$ and $R(a_1)$ is independent of $r$. Hence, the energy spectrum becomes:

$$E_0^{(k)} = \sum_{i=0}^{k} R(a_i) \qquad E_{nl} = E_{nl}^- + E_0 \qquad (28)$$

If we now consider a mapping of the form:

$$B \rightarrow B' = B - a \qquad (29)$$

In Eq. (26), it is easily seen that apart from a constant, the partner potential are the same. In technical words, the chosen SUSY potential satisfies the shape invariance condition. On the other hand, we can obtain:

$$B_1 = B_0 - a \qquad B_n = B_0 - na \qquad (30)$$

$R(a_1) = V_+(B, r) - V_-(B - a, r) =$

$$\frac{\hbar^2}{2\mu}\left[(\frac{-\beta}{2(B^2-\gamma)})^2 + \frac{B^2}{r^2} + \frac{2\left(\frac{-\beta}{2(B^2-\gamma)}\right)B}{r^2} + \frac{B}{r^2} - (\frac{-\beta}{2((B-a)^2-\gamma)})^2 - \frac{(B-a)^2}{r^2} - \frac{2\left(\frac{-\beta}{2((B-a)^2-\gamma)}\right)(B-a)}{r^2} + \frac{B-a}{r^2}\right] \qquad (31)$$

$$\rightarrow R(a_i) = -\frac{\hbar^2}{2\mu}\left[(\frac{-\beta}{2((B-ia)^2-\gamma)})^2 - (\frac{-\beta}{2((B-(i-1))^2-\gamma)})^2\right] \qquad (32)$$

$$R(a_i) = -\frac{\hbar^2}{2\mu}\left[(\frac{-\beta}{2((B-ia)^2-\gamma)})^2 - (\frac{-\beta}{2((B-a(i-1))^2-\gamma)})^2\right] \qquad (33)$$

where the remainder R($a_i$) is independent of $r$. By using Eqs. (24) and (28), the energy levels of the IQY potential are found as:

$$E_{n,l} = -\frac{\hbar^2}{2\mu}\left[\left(\frac{-\beta}{2((B-na)^2-\gamma)}\right)^2 - \left(\frac{-\beta}{2(B^2-\gamma)}\right)^2 + \left(\frac{-\beta}{1+\sqrt{1+4\gamma}}\right)^2\right] + 4a^2V_0 \tag{34}$$

Now we change the potential in Eq. (15) and use the Hulthén potential. Hence this equation changes to:

$$\frac{d^2U_{n,l}}{d^2r} + \frac{2\mu}{\hbar^2}\left(E_{n,l} + \frac{V_0 e^{-\delta r}}{1-e^{-\delta r}} - \frac{\hbar^2(l+6)(l+7)}{2\mu r^2}\right)U_{n,l} = 0 \tag{35}$$

We use the notation $\delta \to -2a$ and Taylor expansion for the Hulthén potential. Now we can introduced the effective potential:

$$V_{eff} = \frac{V_0}{1-e^{2ar}} + \frac{2aV_0 r}{1-e^{2ar}} - \frac{\Omega}{r^2} \tag{36}$$

By using notation $r = \frac{1}{2a}\left(e^{2ar}-1\right)$ [29], the effective potential takes the form of:

$$V_{eff} = \frac{V_0}{1-e^{2ar}} + \frac{2aV_0\left[\frac{1}{2a}(e^{2ar}-1)\right]}{1-e^{2ar}} - \frac{4a^2\Omega}{(e^{2ar}-1)^2} = \frac{V_0}{1-e^{2ar}} - V_0 - \frac{4a^2\Omega}{(1-e^{2ar})^2} \tag{37}$$

By substituting Eq. (37) in Eq. (35), the Schrödinger equation changes to:

$$\frac{d^2U_{n,l}}{d^2r} + \frac{2\mu}{\hbar^2}\left(E_{n,l} - V_0 + \frac{V_0}{1-e^{2ar}} - \frac{4a^2\Omega}{(1-e^{2ar})^2}\right)U_{n,l} = 0 \tag{38}$$

By choosing $\varepsilon_{n,l} = \frac{2\mu}{\hbar^2}(E_{n,l} - V_0)$, $\beta = \frac{4\mu}{\hbar^2}V_0$, $\gamma = \frac{8\mu}{\hbar^2}a^2\Omega$, Eq. (38) takes the form of:

$$\frac{d^2U_{n,l}(r)}{dr^2} + \left(\varepsilon_{n,l} + \frac{\beta}{1-e^{2ar}} - \frac{\gamma}{(1-e^{2ar})^2}\right)U_{n,l}(r) = 0 \tag{39}$$

As we know from SUSY the superpotential defines as:

$$W_1 = -\frac{\hbar}{\sqrt{2\mu}}\left(A + \frac{B}{1-e^{2ar}}\right) \tag{40}$$

If we use Eq. (40) into Ricatti equation, as we did before, gives us $A^2 = -\varepsilon_{n,l}$, $2AB - 2aB = -\beta$, $B^2 + 2aB = \gamma$. So the ground state binding Energy is obtained as:

$$E_{n,l} = \frac{-\hbar^2}{2\mu}\left(\frac{-\beta}{-2a \pm \sqrt{4a^2 + 4\gamma}} + a\right)^2 + V_0 \quad (41)$$

By using Eq. (23) from SUSY method, the ground state normalized eigenfunctions are given as:

$$\psi_0^1(r) = N_0 \exp\left[Ar - \frac{B}{2a}\ln\left((1-e^{2ar})(e^{2ar})\right)\right] \quad (42)$$

In Table 2. the fitted values of parameters of the ground state binding energy equations for Hulthén potential are given.

Table 2. The fitted values of parameters of the ground state binding energy equations for Hulthén potential, column B.E (our model) contains our calculation and the column B.E (experiment) contains the experimental data.

| $a$ (fm$^{-1}$) | $V_0$ (MeV.fm$^2$) | B.E (MeV) | B.E (experiment) [26] |
|---|---|---|---|
| 0.12 | 23 | 31.76 | 31.995 |
| 0.20 | 28 | 31.84 | 31.995 |
| 0.30 | 29 | 30.77 | 31.995 |
| 0.40 | 28 | 28.96 | 31.995 |
| 0.45 | 30 | 30.81 | 31.995 |
| 0.45 | 32 | 32.87 | 31.995 |
| 0.48 | 31.2 | 31.94 | 31.995 |
| 0.48 | 32 | 32.76 | 31.995 |
| 0.48 | 30 | 30.71 | 31.995 |

From Table 2, it can be seen that for $\delta = 0.48$(fm$^{-1}$) and $V_0 = 31.2$(MeV), the calculated ground state binding energy (31.94 MeV) has a good agreement with the experimental data.

As we did before, the energy levels with this potential take the form of:

$$E_{n,l} = -\frac{\hbar^2}{2\mu}\left[\left(\frac{\gamma-\beta}{2(B-na)} - \frac{B-na}{2}\right)^2 - \left(\frac{\gamma-\beta}{2B} - \frac{B}{2}\right)^2 + \left(\frac{-\beta}{-2a + \sqrt{4a^2 + 4\gamma}} + a\right)^2\right] + V_0 \quad (43)$$

## 6. Conclusion

In this paper, we presented a new model for obtaining the binding energy of lithium nucleus for ground state and other as well as the wave functions. For our purpose we solved the redial of Schrödinger equation exact analytically for the IQY potential and the Hulthén potential by using SUSY method. Then, we have presented the formula of wave function for this element. The results obtained from SUSY method for these two potentials are in a good agreement with the experimental data.